\documentclass[trackchanges]{aastex7}

\usepackage{amsmath}
\usepackage{subfig} % 加载 subfig 宏包
\begin{document}

\title{A High-Precision Dynamical Model of Callisto: Incorporating Rotation Effects within Multi-Layer Internal Structure Models}

\author[orcid=0009-0001-4162-2792]{Kai Huang}
%\altaffiliation{University of Chinese Academy of Science, Beijing 100049, China}
\affiliation{Yunnan Observatories, Chinese Academy of Sciences, Kunming 650216, China}
\affiliation{University of Chinese Academy of Science, Beijing 100049, China}
\email{huangkai@ynao.ac.cn}  

\author[orcid=0000-0001-9202-7750]{Yongzhang Yang} 
%\altaffiliation{Las Campanas Observatory}
\affiliation{Yunnan Observatories, Chinese Academy of Sciences, Kunming 650216, China}
\email{yang.yongzhang@ynao.ac.cn}

\author[0000-0002-8077-094X]{Yuhao Chen}
%\altaffiliation{Astrosat Post-Doctoral Fellow}
\affiliation{Yunnan Observatories, Chinese Academy of Sciences, Kunming 650216, China}
\affiliation{University of Chinese Academy of Science, Beijing 100049, China}
\email{chenyuhao@ynao.ac.cn}

\author[0000-0001-5933-5794]{Yining Zhang}
\affiliation{University of Chinese Academy of Science, Beijing 100049, China}
\affiliation{National Astronomical Observatories, Chinese Academy of Sciences,Beijing 100101, China}
\email{zhangyn@bao.ac.cn}

\author[0000-0003-1022-8423]{Yuqiang Li}
\affiliation{Yunnan Observatories, Chinese Academy of Sciences, Kunming 650216, China}
\affiliation{Key Laboratory of Space Object and Debris Observation, PMO, CAS, Nanjing 210023, China}
\email{lyq@ynao.ac.cn}

\correspondingauthor{Yongzhang Yang} 
\email{yang.yongzhang@ynao.ac.cn}
\correspondingauthor{Yuqiang Li} 
\email{lyq@ynao.ac.cn}
%\collaboration{all}{The Terra Mater collaboration}

\begin{abstract}
	
	China is planing to launch the “Tianwen-4” mission around the year 2030, with its aim being the exploration of Jupiter and its moon, Callisto. Within the realm of deep space exploration, the accuracy of ephemerides is of great importance. Current ephemerides employ a simplified rotation model for Callisto, which this study addresses by proposing a novel dynamical model. This model enhances the existing orbital dynamics by integrating Callisto’s rotational motions influenced by gravitational torques from the Sun, Jupiter, and other Galilean moons within an inertial frame, capturing the intricate coupling between Callisto’s orbital and rotational dynamics. The study establishes a full dynamical model by deriving analytical expressions for this coupling and developing an adjustment model for data fitting using precise orbit determination methods. Furthermore, the influence of tidal effects on Callisto’s motion is investigated, considering its multi-layered internal structure. Results demonstrate that the difference between the newly established full model and the model in  current ephemerides is on the order of tens of meters. When calculating the impact of different internal structures of Callisto on its orbit, the influence of three-layered and two-layered structures is on the order of meters, suggesting that the development of a high-precision dynamical model requires additional constraints on the internal structure of Callisto. This research provides a novel alternative for a new generation of precise numerical ephemerides for Callisto. Additionally, these findings provide a testing platform for the data from the “Tianwen-4” mission.
	
\end{abstract}



\section{Introduction}
Jupiter is the largest planet in the Solar System and possesses more than 90 moons. The Jupiter system is regarded as a microcosm of the Solar System, comprising four Galilean satellites—Io, Europa, Ganymede, and Callisto—as well as several smaller moons. The study of Galilean satellites is of significant importance for understanding the formation and evolution of the Jupiter system and thus the Solar System, making the exploration of the Jupiter system a key scientific theme for future space missions \citep{Genova2022GeodeticIO}.

In recent years, humanity has launched numerous spacecraft missions to the Jupiter system, including flybys and orbits. These missions encompassed the Pioneers 10 and 11, Voyagers 1 and 2, Ulysses, Galileo, Cassini, New Horizons, and Juno~\citep{Wang2023JupitersCS}. The Jupiter Icy Moons Explorer (JUICE), a mission by the European Space Agency, was launched on April 14, 2023, and is expected to arrive at Jupiter and its moons in July 2031~\citep{Grasset2013JUpiterIM,Elsaesser2023FutureSE,Poulet2024MoonsAJ}. China plans to launch the Tianwen-4 mission to explore Jupiter and Callisto around 2030~\citep{Lin2018ChinasPF}. The fundamental physical parameters of Jupiter and its four Galilean satellites have been thoroughly studied in previous exploration activities, such as the gravitational constant, gravitational field, and the tidal factor $k_{2}$ of the Io~\citep{Iess2018MeasurementOJ,Durante2020JupitersGF,VanHoolst2024GeophysicalCO,Tosi2024CharacterizationOT,Sun2024AssessmentOC,Park2024IosTR}. Based on these studies, the dynamical models of the Galilean moons and their ephemerides have become feasible.

The precise determination of the internal structure of the Galilean satellites is crucial for a deeper understanding of the formation mechanism of the Jovian system and the evolutionary history of its moons. Unlike Europa, Io, and Ganymede, which have undergone significant geological evolution, Callisto’s surface has remained largely unaltered by internal geological activity since its formation 4.5 billion years ago ~\citep*{Kuskov2005,Greeley2000GalileoVO}. As indicated by its moment of inertia, Callisto’s internal 
structure is inferred to be partially differentiated, corresponding to a scenario 
of incomplete separation of rock and ice ~\citep{Anderson2001ShapeMR,VanHoolst2024GeophysicalCO}. Therefore, Callisto has become the best preserver of its formation process and ancient history. 
Callisto also lacks an intrinsic magnetic field~\citep{Khurana1998InducedMF}, but its response to changes in Jupiter’s magnetic field is similar to the electrical conductivity of water in the ocean at depths less than about 200 kilometers~\citep{Zimmer2000SubsurfaceOO}. Therefore, scientists infer that there is a liquid ocean on Callisto ranging from 20 to 200 kilometers in depth. It is precisely because of these characteristics that Callisto has become a key target for exploration in the Jovian system. In order to gain a deeper understanding of Callisto, China plans to launch the “Tianwen-4” spacecraft around 2030 for a close-up exploration of Callisto. This mission is of great significance for us to understand Callisto’s gravitational field, internal structure, atmosphere, and other aspects~\citep{Xu2022ABI}. In this paper, a new dynamical model of Callisto will be introduced, which will serve as the foundational dynamical model for producing the next generation of high-precision ephemerides for the Galilean moons, and will provide a valuable reference for the orbital calculations of the Tianwen-4 mission.

The first ephemerides of the Galilean moons are constructed using analytical methods, primarily based on the Sampson-Lieske theory. The dynamical model was initially developed early last century by \citet{Sampson1920OnCE}, and subsequently refined by \citet{Lieske1977ExpressionsFT} to meet the requirements of the Voyager spacecraft mission. The theory was further developed by \citet{Giorgini1996JPLsOS}, and this advancement is known as the E5. \citet{Lainey2004NewAE2,Lainey2004NewAE} further refined the model using numerical methods, which includes a two-body model of Callisto’s orbit around Jupiter, Jupiter’s gravitational field, a third-body perturbation model of the major bodies of the solar system, the gravitational fields of the three Galilean satelliites ,general relativistic effects and the effects of physical librations. On the basis of this model, several new numerical ephemerides for the Galilean moons have been developed~\citep{Kosmodamianskii2009NumericalTO,Kosmodamianskiy2019UpdatedNE}. These ephemerides adopt a dynamical model that includes a simplified rotation model of Callisto.

This paper presents a new high-precision dynamical model that includes the rotation of Callisto. The research on high-precision dynamical models of the Galilean satellites has two main implications. On one hand, as the precision of spacecraft observations of the satellites improves, we require ephemerides of high precision, and the accuracy of the dynamical model greatly affects the precision of the generated ephemerides. On the other hand, different internal structures will result in different rotations of the Galilean satellites when subjected to torques. Conversely, we can infer the internal structures of the satellites from their rotational states. Studying the motion of the satellites through dynamical models is a primary method for investigating the internal structures of the Galilean satellites. Therefore, this study builds upon existing dynamical models of the Galilean satellites, referencing the methods of rotation models for the Moon and Martian moons~\citep{Rambaux2012RotationalMO,Folkner2014ThePA,Pavlov2016DeterminingPO,Yang2020AnEM,Yang2024NumericalMO,articleHUANG}, and considers the torques exerted by the Sun, Jupiter, and the three Galilean satellites on Callisto. It investigates the rotation of Callisto in International Celestial Reference System (ICRS). Based on the generated rotation model, the gravitational field of Callisto is incorporated into the dynamical model of its orbital motion to study the coupling effects between Callisto’s rotation and orbital motion. A new dynamical model describing the motions of Callisto is established. 

The data fitting work in this paper will be divided into two parts:
(1) Fit the simple model to the position and velocity data of Callisto in the NOE-5-2023 ephemeris. Through this method, the parameters of the reproduced dynamical model are optimized, and the fitting data are provided for the establishment of the full model.
(2) Establish a dynamical model coupling rotation and orbit, use the fitting orbit data obtained in step (1) to perform data adjustment on the new model, and then analyze the differences between the two modeling methods.

Additionally, based on the established full dynamical model, we further considered the tidal effects of Callisto within the model. According to the modeling of Callisto’s internal structure in existing research, Callisto may have a double-layer structure or a three-layer model consisting of an ice shell, water, and rock~\citep{Anderson2001ShapeMR,Nagel2003AMF,Kuskov2005InternalSO}. We simulated the precise orbit determination results of tidal effects under these two structures within the full dynamical model.

The structure of this paper is organized as follows: At first, the establishment of the current dynamical model of Callisto and the complete dynamical model developed in this study is introduced, along with a quantitative analysis of the impact of rotational effects on the dynamical model. Subsequently, the method and conclusions for establishing the least squares adjustment model for data fitting of the developed complete dynamical model using precise orbit determination are described. Finally, the influence of tidal effects due to different internal structures of Callisto within the model is explored.

\section{Establishment of Callisto's Dynamical Model in the Current Ephemerides}

In order to clarify the difference between the dynamical model used in the current ephemerides and the dynamical model to be established in this paper, which includes Callisto’s rotation, this section will reproduce the dynamical model from the current ephemerides. Then, in the next section we will incorporate Callisto’s rotation to establish a dynamical model that couples the Callisto’s rotation with its orbital motion.

\subsection{Reproduce the dynamical model used in ephemerides}
In this paper, we refer to the dynamical model in current ephemerides as the simple model, and the model including rotation as the full model. In this section, we will follow the research method of \citet{Lainey2004NewAE2,Lainey2004NewAE,Lainey2019Interior} to establish a simple model describing Callisto’s orbital motion in the ICRS. For natural satellites, the gravity of the central body is the primary source of their dynamical force, and their equations of motion can be expressed as equation~(\ref{eq1}):
\begin{equation}
\frac{d^2 \boldsymbol{r}}{d t^2}=-\mu \frac{\boldsymbol{r}}{r^3}+\frac{\boldsymbol{F}}{m}.
\label{eq1}
\end{equation}
$\boldsymbol{r}$ represents the position vector between Callisto and Jupiter; $r$ is the distance between Callisto and Jupiter; $\mu = G(M + m) $, $G$ is the gravitational constant, $M$ and $m$ are the masses of Jupiter and Callisto, respectively; the first term $-\mu \frac{\boldsymbol{r}}{r^3}$ is the gravitational force of the central body; $\boldsymbol{F}$ is the sum of the other perturbing forces, which mainly include the non-spherical perturbations of the central body and the three Galilean satellites, the N-body perturbations of all
planets and Pluto and the three Galilean satellites, and general relativistic effects. In the work of \citet{Lainey2019Interior}, this dynamical model of Galilean satellites is summarized in the following formula~(\ref{eq2}):
\begin{equation}
\begin{aligned}
	\ddot{\boldsymbol{r}}= & -\frac{G\left(m_0+m\right) \boldsymbol{r}}{r^3}+\sum_{j=1, j \neq i}^{\mathcal{N}} G m_j\left(\frac{\boldsymbol{r}_j-\boldsymbol{r}}{r_{i j}^3}-\frac{\boldsymbol{r}_j}{r_j^3}\right) \\
	& +G\left(m_0+m\right) \nabla_i U_{\hat{\imath} \hat{0}}+\sum_{j=1, j \neq i}^{\mathcal{N}} G m_j \nabla_j U_{\hat{\jmath} \hat{0}} + \boldsymbol{a}_{rel}+ \boldsymbol{a}_{lib} .
\end{aligned}
\label{eq2}
\end{equation}
In Equation~(\ref{eq2}), the first term on the right-hand side represents the two-body force between Callisto and Jupiter, the second term is the third-body perturbation force from other celestial bodies in the solar system, the third term accounts for the gravitational field effect of Jupiter, the fourth term represents the influence of the gravitational fields of other celestial bodies,  the fifth term is the effect of general relativity, whose form is given in section~\ref{genrel}, and the last term is the effect of physical librations.

In the process of constructing the dynamical model, this paper will use this model to complete the establishment of a simple dynamical model for Callisto. Then, a new modeling method will be proposed to model Callisto’s rotation, and the gravitational field of Callisto will be introduced into the dynamical model, completing the establishment of a full dynamical model that couples rotation with orbital motion.

\begin{table}[!htp]
\centering
\caption{\textcolor{black}{The gravitational constant of celestial bodies.}}\vspace{1mm}
\label{table1}
\begin{tabular}{lc}
	\hline celestial bodies  & $\mathrm{GM}(\mathrm{km}^3 / \mathrm{s}^2)$ \\
	\hline Sun  & 132712440041.939400 \\
	Jupiter  &  126686595.050296\\
	Io & 5949.29470829565 \\
	Europa  & 3201.07755988779 \\
	Ganymede & 9910.52558574655 \\
	Callisto & 7103.31440370675 \\
	\hline
\end{tabular}
\end{table}

\subsection{N-body force}

Beyond the gravitational attraction exerted by the central body, Jupiter, Callisto is further influenced by the gravitational forces of other bodies within the solar system. This study takes into account the impacts of the Sun, all planets, and the three Galilean satellites. The values of the gravitational constant GM are presented in Table~\ref{table1}, and the gravitational constant of celestial bodies in the context of the NOE ephemerides. The ephemeris file for the Galilean satellites used in this paper is NOE-5-2023, which is provided by IMCCE and contains positional information of the Galilean satellites from 1879 to 2049. The model was fit to all Earth-based (including radar) observations from 1891 to 2022, as well as HST, Voyager, Galileo imaging data and Gaia data. In the ICRS, the position vectors for the Jupiter barycenter, from Jupiter to Callisto, and the perturbing body j are denoted as $\boldsymbol{r}_m$, $\boldsymbol{r}$, and $\boldsymbol{r}_j$, respectively. The acceleration of Callisto is given by eqution~(\ref{eq3}):
\begin{equation}
\label{eq3}
\boldsymbol{a}=-\frac{G M_{\text {jupiter }}}{r^2} \frac{\boldsymbol{r}}{r}-\frac{G M_j}{\Delta_j^2} \frac{\Delta_{\boldsymbol{j}}}{\Delta_j},
\end{equation}
\begin{equation}
\Delta_j=\boldsymbol{r}-\boldsymbol{r_j}.
\end{equation}

The equation~\eqref{eq3} represents the acceleration of Callisto relative to the ICRS. However, when studying the motion of Callisto, its movement must be described within the Jovian system. So we need to convert this acceleration in ICRS to the Jupiter-centered Clestial Reference System(JCRS). The JCRS maintains the same axis orientations as the ICRS, however, its reference point is located at the center of Jupiter.
\begin{equation}
\label{eq4}
\boldsymbol{a}=-\frac{G M_{\text {jupiter }}}{r^2} \frac{\boldsymbol{r}}{r}- \ G M_j \left(\frac{\boldsymbol{r}_{\boldsymbol{j}}}{r_j{ }^3}+\frac{\Delta_{\boldsymbol{j}}}{\Delta_j^3}\right).
\end{equation}

The first term in the formula~(\ref{eq4}) represents the two-body acceleration between Jupiter and Callisto, while the acceleration generated by the perturbing satellite Callisto is the second term. The planetary positions and main physical parameters used in this model are sourced from the ephemerides NOE-5-2023.

\subsection{Jupiter’s Gravitational Field and Jupiter’s Rotation}
Due to Jupiter’s irregular shape and non-uniform 
internal mass distribution, it cannot be treated as an ideal spherical body when calculating the gravitational force on its satellites. Non-spherical perturbations must be taken into account. During the motion of Callisto, the coordinate system employed is the JCRS, whereas the gravitational field is defined in the Jupiter body-fixed frame. Therefore, calculating the acceleration generated by Jupiter’s gravitational field on Callisto requires an appropriate rotation model to facilitate the transformation between the two coordinate systems. There has been extensive research on Jupiter’s coordinate systems~\citep{Simon2013NewAP,Bagenal2016SurveyOG,Connerney2018ANM,Archinal2018ReportOT,Connerney2021ANM}, In these studies, the body-fixed coordinate system of Jupiter is usually defined based on the magnetic field characteristics of Jupiter. As for our work, it is based on the research conducted by~\citet{Archinal2018ReportOT}, which depict the relationship between the Jupiter-fixed coordinate system and the inertial frame as illustrated in Figure~\ref{Fig1}. The calculation formula for converting the position $\boldsymbol{R}_{B F}$ in the Jupiter-fixed coordinate system to the position $\boldsymbol{R}_{I N}$ in the JCRS is as eqution~(\ref{eq6}):
\begin{figure}
\centering
\includegraphics[width=0.6\hsize]{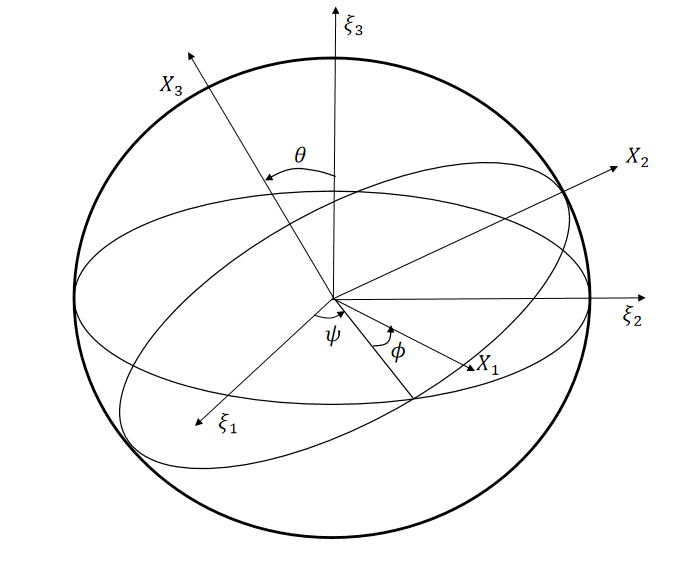}
\caption{The Euler angles for the transformation between JCRS and the Jupiter-fixed coordinate system~\citep{Yang2017DeterminationOT}.
	Precession angle($\phi$): The angle at which Jupiter rotates around the Z-axis of the fixed coordinate system;
	Nutation angle($\theta$): The angle at which Jupiter rotates around the X-axis obtained after the first rotation (the intersection line of Jupiter's celestial coordinate system and the equatorial plane of the fixed coordinate system);
	Spin angle($\psi$): The angle at which the rigid body rotates around the Z-axis obtained after the second rotation.}
\label{Fig1}
\end{figure}
\begin{equation}
\boldsymbol{R}_{I N}=R_Z(-\psi) R_X(-\theta) R_Z(-\phi) \boldsymbol{R}_{B F}.
\label{eq6}
\end{equation}
where $\psi,\theta,\phi$ are the Euler angles describing the rotation of Jupiter. Jupiter's Euler angles describe the orientation of its rotational axis within an inertial reference frame. Specifically, these angles include the precession angle (denoting the rotation of the spin axis about the ecliptic axis), the nutation angle (representing the inclination between the spin axis and the ecliptic axis), and the spin angle (indicating the rotation about the spin axis itself), collectively characterizing variations in its spatial orientation.
$R_X,R_Z$ are rotation matrices 
about Jupiter’s X and Z body-fixed axes respectively.

We calculate the values of the three Euler angles at any instant of time according to the method of~\citet{Archinal2018ReportOT}.
\begin{equation}
\begin{aligned}
	\alpha_0= & 268.056595-0.006499 T+0^{\circ} .000117 \sin \mathrm{Ja}+0^{\circ} .000938 \sin \mathrm{Jb} \\
	& +0.001432 \sin \mathrm{Jc}+0.000030 \sin \mathrm{Jd}+0.002150 \sin \mathrm{Je} \\
	\delta_0= & 64.495303+0.002413 T+0.000050 \cos \mathrm{Ja}+0.000404 \cos \mathrm{Jb} \\
	& +0.000617 \cos \mathrm{Jc}-0.000013 \cos \mathrm{Jd}+0.000926 \cos \mathrm{Je} \\
	W= & 284.95+870.5360000 d^{(\mathrm{e})} \\
\end{aligned}
\end{equation}
\begin{equation}
\begin{aligned}
	\mathrm{Ja}= & 99^{\circ} .360714+4850^{\circ} .4046 T, \quad \mathrm{Jb}=175^{\circ} .895369+1191^{\circ} .9605 T, \\
	\mathrm{Jc}= & 300^{\circ} .323162+262^{\circ} .5475 T, \quad \mathrm{Jd}=114^{\circ} .012305+6070^{\circ} .2476 T, \\
	\mathrm{Je}= & 49^{\circ} .511251+64^{\circ} .3000 T
\end{aligned}
\end{equation}
\begin{equation}
\begin{aligned}
	\phi & =\alpha_0+90^{\circ} \\
	\theta & =90^{\circ}-\delta_0 \\
	\psi & = W
\end{aligned}
\end{equation}
$T$ is the Interval in Julian centuries (36,525 days) from the epoch J2000.0 and $d$ is the Interval in days from the epoch J2000.0.

The gravitational potential function is expanded into spherical harmonics as follows:
\begin{equation}
V=\frac{G M}{r} \sum_{n=2}^{\infty} \sum_{m=0}^n\left(\frac{R_{\text {Jupiter }}}{r}\right)^n \bar{P}_{n m}(\sin \phi)\left[\bar{C}_{n m} \cos m \lambda+\bar{S}_{n m} \sin m \lambda\right].
\end{equation}
In this expression, $GM$ is the gravitational constant of Jupiter, $R_{\text {Jupiter }}$ is the radius of Jupiter, $\bar{P}_{n m}$ represents the normalized associated Legendre function, $\bar{C}_{n m}$ and $\bar{S}_{n m}$ are the normalized spherical harmonic coefficients of the Jupiter gravity field model. $\lambda$, $\phi$ and $r$ denote the longitude, latitude and distance to the center of Jupiter in the fixed coordinate system of Callisto on Jupiter. Much research has been conducted on Jupiter’s gravitational field based on multiple Jupiter exploration missions~\citep{Iess2018MeasurementOJ,Durante2020JupitersGF}. In this paper, we utilize the gravitational field model developed by~\citet{Iess2018MeasurementOJ}, with its coefficients listed in Table~\ref{table11}.
It is necessary to follow the coordinate transformation method described in the previous section. First, the coordinates of Callisto in the JCRS should be converted to the Jupiter-fixed frame. Subsequently, the integration calculations are performed by transforming from the Jupiter-fixed frame back to the JCRS.
\begin{table}
\centering
\caption{Unnormalized gravity field coefficients of Jupiter.}\vspace{1mm}
\label{table11}
\begin{tabular}{lrr}
	\hline & \multicolumn{1}{c}{ Value } & Uncertainty \\
	\hline$J_2\left(\times 10^{-6}\right)$ & $14,696.572$ & 0.014 \\
	$C_{21}\left(\times 10^{-6}\right)$ & -0.013 & 0.015 \\
	$S_{21}\left(\times 10^{-6}\right)$ & -0.003 & 0.026 \\
	$C_{22}\left(\times 10^{-6}\right)$ & 0.000 & 0.008 \\
	$S_{22}\left(\times 10^{-6}\right)$ & 0.000 & 0.011 \\
	$J_3\left(\times 10^{-6}\right)$ & -0.042 & 0.010 \\
	$J_4\left(\times 10^{-6}\right)$ & -586.609 & 0.004 \\
	$J_5\left(\times 10^{-6}\right)$ & -0.069 & 0.008 \\
	$J_6\left(\times 10^{-6}\right)$ & 34.198 & 0.009 \\
	$J_7\left(\times 10^{-6}\right)$ & 0.124 & 0.017 \\
	$J_8\left(\times 10^{-6}\right)$ & -2.426 & 0.025 \\
	$J_9\left(\times 10^{-6}\right)$ & -0.106 & 0.044 \\
	$J_{10}\left(\times 10^{-6}\right)$ & 0.172 & 0.069 \\
	$J_{11}\left(\times 10^{-6}\right)$ & 0.033 & 0.112 \\
	$J_{12}\left(\times 10^{-6}\right)$ & 0.047 & 0.178 \\
	\hline
\end{tabular}
\end{table}

In calculating the acceleration of the gravitational field, a recursive approach is commonly used to avoid complex differentiation processes. In this study, the~\citet{Cunningham1970OnTC} method is used to solve for the acceleration of Jupiter’s gravitational field.

\subsection{Galilean satellites gravitational field}
In the Jupiter system, due to the large mass of the Galilean satellites, they exert a strong gravitational influence on Callisto. We thus consider the gravitational field of the Galilean satellites’(Io, Europa, Ganymede) influence on Callisto. The data of the Galilean satellites’ gravitational fields are listed in Table \ref{tabel3}\citep{Anderson2001ShapeMR,Jacobson1999ACO,Kloster2011EuropaOT,Casajus2020UpdatedEG}. Meanwhile, the rotational model used is the IAU rotational model~\citep{Archinal2018ReportOT}. The International Astronomical Union (IAU) revises the recommended values for coordinate systems and rotational parameters of planets, satellites, asteroids, and comets in the solar system. According to IAU standards, the coordinate systems of planets and satellites are established based on their mean rotation axes, with spatial orientation achieved through specific longitude definitions. For celestial bodies with observable rigid surfaces, their longitude systems typically use surface features (such as impact craters) as reference datums. IAU proposed analytical expressions that correlate celestial rotational parameters with the ICRF. For the work in this paper, when studying Callisto, we will perform synchronous integration calculations of its orbit and rotation to investigate the coupled effect of rotation on the orbit. Therefore, we will establish a motion model of Callisto from a dynamical perspective instead of using the IAU rotational model.

\begin{table}
\centering
\caption{Gravitational Harmonics of Galilean satellites (units of $10^{-6}$)}\vspace{1mm}
\label{tabel3}
\begin{tabular}{cccccc}
	\hline Satellite & $J_2$ & $C_{21}$ & $S_{21}$ & $C_{22}$ & $S_{22}$ \\
	\hline  Io  & 1854.9 & 0& 0 &553 & 0 \\
	Europa  & 461.39 & 4.25& 0 &138.42 & 0 \\
	Ganymede  & 240 & 0& 0 &72.1 & 0 \\
	Callisto  & 32.7 & 0& 0 &10.2 & -1.1 \\
	\hline
\end{tabular}
\end{table}

\begin{figure}
\centering
\includegraphics[width=0.8\hsize]{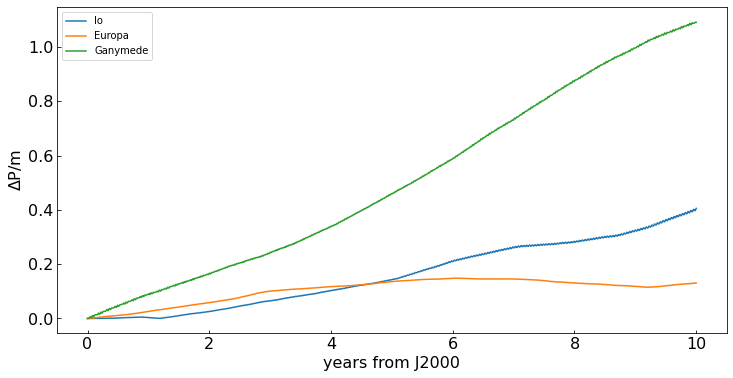}
\caption{The influence of the gravitational fields of the three Galilean satellites (Io, Europa, and Ganymede) on Callisto’s orbit. The ordinate $\mathit{\Delta}P$  represents the difference between the orbit that includes the perturbations of the Galilean satellites and the calculated orbit that does not, with the unit being meters; the abscissa represents the integration duration starting from the J2000 epoch.}
\label{Figure3}
\end{figure}

After introducing the gravitational fields of the three Galilean satellites, the impact on the dynamical model of Callisto is illustrated in Figure~\ref{Figure3}. The gravitational fields of the other Galilean moons cause variations in Callisto's orbit on the order of meters. 
The magnitude of this effect is relatively small compared to other physical effects influencing the orbital motion of Callisto, and it can be neglected during flyby or orbital missions of artificial satellites around Callisto. However, it must be considered when executing landing exploration missions.

\subsection{physical librations}

The gravitational influence of surrounding celestial bodies on Callisto can lead to an imperfect synchronization between Callisto’s rotation and orbit, resulting in librations effects. In the model established in this paper, we refer to the completed research on physical librations\citep{1990Phobos,Jacobson2010The,Lainey2019Interior}:
\begin{equation}
\Delta \varpi=\frac{3}{2}\left(\frac{R}{a}\right)^2\left[J_2-2 C_{22}\left(5-\frac{4 \theta}{e}\right)\right] n t+\frac{3}{2}\left(\frac{R}{a}\right)^2 \frac{\left(J_2+6 C_{22}\right)}{e} \sin (M)
\end{equation}
The parameters $a$, $e$, $n$, $M$, and $\varpi$ represent the conventional osculating Keplerian elements: semimajor axis, eccentricity, mean motion, and mean anomaly, respectively. $R$, $J_2$ and $C_{22}$ refer to the mean radius and unnormalized gravitational coefficients of Callisto. This formulation explicitly demonstrates the influence of the libration amplitude $\theta$ on the periapsis. 

\subsection{General Relativity perturbation model}
\label{genrel}
During the motion of Callisto, it is mainly influenced by the General Relativity effect generated by the mass of Jupiter. the acceleration induced by the General Relativity effect of Jupiter on Callisto is given by:
\begin{equation}
\boldsymbol{a}_{r e l}=-\frac{G M}{c^2 r^3}\left\{\left[4 \frac{G M}{r}-\boldsymbol{v}^2\right] \boldsymbol{r}+4(\boldsymbol{r} \cdot \boldsymbol{v}) \boldsymbol{v}\right\}
\end{equation}
Where $GM$ is the gravitational constant of Jupiter, 
$\boldsymbol{v}$ is the velocity of Callisto in the JCRS, 
$\boldsymbol{r}$ is the position vector from the jupiter to Callisto, and 
$c$ is the speed of light.

\subsection{The result of simple model}

\begin{figure}
\centering
\includegraphics[width=0.8\hsize]{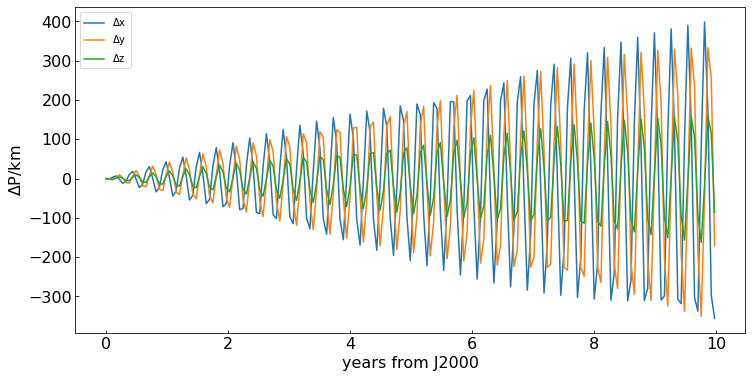}
\caption{The integration starts at the J2000 epoch. By reproducing the dynamical model from the current ephemeris, we calculate and compare the orbital differences with the position of Callisto from NOE-5-2023 ephemeris. The ordinate shows positional differences (in kilometers), and the abscissa represents the integration duration (in years).}
\label{Figure4}
\end{figure}

\begin{figure}
\centering
\includegraphics[width=0.8\hsize]{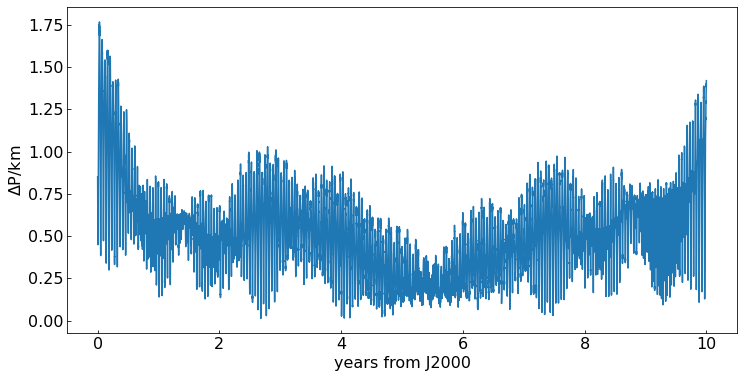}
\caption{The integration commences at the J2000 epoch. The reproduced dynamical model is subjected to data fitting using position and velocity data of Callisto from the NOE-5-2023 ephemeris. The ordinate depicts positional differences (in kilometers), while the abscissa represents the integration duration (in years).}
\label{simple_fit}
\end{figure}

\begin{figure}
\centering
\includegraphics[width=0.8\hsize]{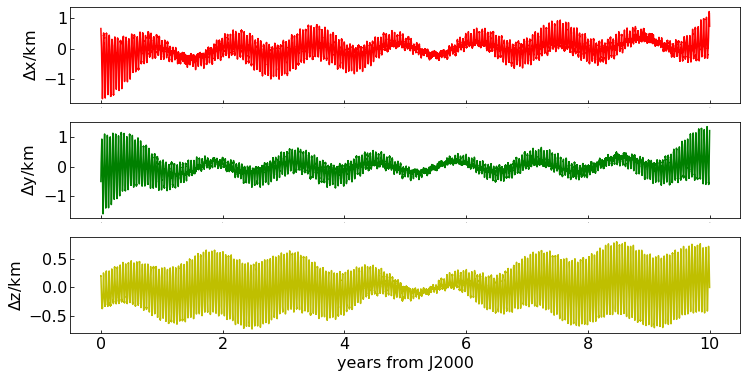}
\caption{After fitting the reproduced model of Callisto to the NOE-5-2023 ephemeris, the differences in the x, y, and z directions between the calculated results and the positions of Callisto in the NOE-5-2023 ephemeris.}
\label{simple_fitxyz}
\end{figure}

We have established a dynamical model similar to the current ephemeris (simple model). By integrating the simple model using the ABM algorithm, we obtained the difference from the NOE-5-2023 ephemeris as shown in Figure~\ref{Figure4}. The difference between the new model and the NOE-5-2023 ephemeris increases linearly, with the maximum difference being 400 km. 
For this study, this represents a significant discrepancy. We will therefore apply precision orbit determination techniques to mitigate the deviations between our newly developed model's computational results and the NOE-5-2023 ephemeris. Then, we used the least squares method and utilized the Callisto data from the NOE-5-2023 ephemeris to fit the new model, and the results are shown in Figure~\ref{simple_fit} and Figure~\ref{simple_fitxyz}. The discrepancy between the two is significantly reduced. Since we cannot determine the exact physical parameters used in the NOE-5-2023 ephemeris dynamical model, the physical parameters of our reconstructed model do not completely match those of the NOE-5-2023 ephemeris. This results in significant differences between Callisto’s orbit and the NOE-5-2023 ephemeris before fitting.
After optimizing Callisto’s initial state parameters through data fitting, these differences are significantly reduced. This indicates that the orbital discrepancies caused by differing physical parameters between the two models are on the order of kilometers, which falls within the acceptable accuracy range for current Callisto ephemeris requirements.
This reconstruction of the simple model also establishes a clear baseline for comparison with the full model—which incorporates rotational effects—to be developed subsequently in this study.

\section{The Establishment of a Full Model Coupling Rotation and Orbit.}

The influence of a natural satellite's rotation on its orbital motion is mainly reflected in the following two aspects: First, the rotational motion of the natural satellite will cause the time-varying characteristics of its own gravitational field in the inertial reference frame. This non-spherical gravity will interact with the gravity of the primary planet, resulting in periodic orbital perturbations. For example, the precession of the rotation axis will cause the orbital plane to rotate slowly, which needs to be corrected in the ephemeris dynamical model. In addition, the coupling between rotation and orbit is also reflected in the tidal lag effect. The difference between the satellite's rotation speed and orbital speed will lead to deformation lag, generating a torque that causes the orbit to gradually expand. This effect needs to be considered in long-term ephemeris predictions.

\subsection{Rotation Model}
In this section, we will incorporate Callisto’s rotation in JCRS and its gravitational field into the simple model to establish a full model describing Callisto’s motion. It should be noted that this study assumes zero obliquity for Callisto in establishing its rotational model. We first establish Callisto’s rotation model under the torque action of Jupiter, the Sun, and the three Galilean satellites based on the rigid body rotation theory. Based on this rotation model, we can calculate the force exerted by Callisto’s gravitational field on other celestial bodies at any given moment, as well as the acceleration caused by the reaction force received by Callisto. And we can establish a dynamical model that couples Callisto’s rotation with its orbital motion. The gravitational field coefficients of Callisto are listed in Table~\ref{tabel3}.

rigid body refers to an idealized model where the shape and size remain unchanged, and the distances between all its particles are always constant (ignoring the deformation of actual objects). This simplification makes it a fundamental tool for analyzing the rotational and translational motions of objects. The differential equations describing the rotation of a rigid body are the Euler-Liouville equations , which form the basis for establishing Callisto’s rotation model. It can be expressed in an inertial frame as:
\begin{equation}
\label{eq18}
\frac{d(I \boldsymbol{\omega})}{d t}+\boldsymbol{\omega} \times I \boldsymbol{\omega}=\boldsymbol{N}.
\end{equation}
\begin{equation}
I=\left(\begin{array}{lll}
	A & 0 & 0 \\
	0 & B & 0 \\
	0 & 0 & C
\end{array}\right).
\end{equation}
$\boldsymbol{\omega}$ represents the angular velocity of the rigid body Callisto in JCRS, $\boldsymbol{N}$ is the torque acting on Callisto (in this study, the sources of the torques discussed are the Sun, Jupiter, and the three Galilean satellites), and $I$ is the moment of inertia tensor matrix of Callisto. When describing the rotation of Callisto, we take the center of mass of Callisto as the origin and the three principal axes of Callisto’s inertia as the three coordinate axes to establish a principal axis (PA) reference frame that rotates with Callisto. The transformation between the PA of Callisto and the Callisto-centered Celestial Reference System(CCRS) is depicted in Formula~(\ref{eq188}).
\begin{equation}
\label{eq188}
\boldsymbol{R}_{CCRS}=R_Z(-\phi) R_X(-\theta) R_Z(-\psi) \boldsymbol{R}_{PA}.
\end{equation}
$\psi,\theta,\phi$ are the Euler angles describing the rotation of Callisto. The formula for 
$\boldsymbol{\omega}$ expressed in terms of the three Euler angles and their rates of change is:
\begin{equation}
\label{eq19}
\begin{gathered}
	\omega_x=\dot{\phi} \sin \theta \sin \psi+\dot{\theta} \cos \psi \\
	\omega_y=\dot{\phi} \sin \theta \cos \psi-\dot{\theta} \sin \psi \\
	\omega_z=\dot{\phi} \cos \theta+\dot{\psi}
\end{gathered}
\end{equation}

However, when employing the equation~\eqref{eq19}, one may encounter singular situations. In such cases, using Wisdom angles instead of Euler angles might be more beneficial~\citep{Wisdom1984TheCR}. 

The value of the angular acceleration $\dot{\boldsymbol{\omega}}$ depends on the torque 
$N$ applied to Callisto and the moment of inertia $I$ of Callisto and the value is 0.3549~\citep{Anderson2001ShapeMR}, which is derived assuming Callisto is in hydrostatic 
equilibrium. The expression for $\dot{\boldsymbol{\omega}}$:
\begin{equation}
\dot{\boldsymbol{\omega}}=I^{-1}(\boldsymbol{N}-\boldsymbol{\omega} \times I \boldsymbol{\omega})
\end{equation}

Analyzing the torques acting on Callisto, this paper will consider the following components of the torque acting on Callisto in inertial space: (1) Within the Callisto gravitational field, treating Jupiter as a point mass to generate a torque on Callisto; (2) Treating the Sun as a point mass, the torque on Callisto generated by the Sun’s point mass due to the presence of Callisto’s gravitational field; (3) Within the Callisto gravitational field, treating Io, Europa, and Ganymede as point masses to generate torques on Callisto, respectively; (4) Due to Jupiter’s irregular shape, under the influence of Callisto’s gravitational field, the torque on Callisto generated by Jupiter’s J2 term~\citep{Breedlove1977ANS}. The formula representing these torques can be expressed as:
\begin{equation}
\label{eq22}
\boldsymbol{N}=\boldsymbol{N}_{\oplus}+\boldsymbol{N}_{\odot}+\boldsymbol{N}_{j}+\boldsymbol{N}_{\oplus \mathbf{F}}
\end{equation}
The solution for each term in the formula can be found in \citet{articleHUANG}.

\begin{figure}
\centering
\includegraphics[width=0.8\hsize]{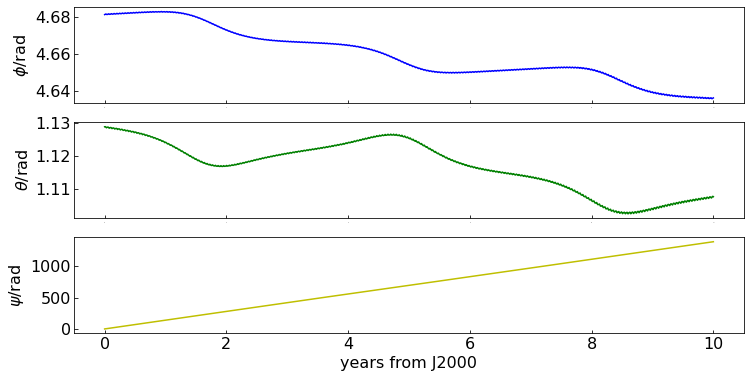}
\caption{The established rotation model of Callisto integrated, and the temporal variations of the three Euler angles are presented.}
\label{Figure5}
\end{figure}

In this study, the Callisto’s motion is taken at the J2000.0 epoch. According to the established model, the differential equations are integrated using the ABM (Adams-Bashforth-Moulton) numerical integration algorithm. The results of the three Euler angles are shown in Figure~\ref{Figure5}. At this point, we have completed the modeling of Callisto's rotation. The obtained results will be utilized to incorporate the influence of Callisto's own gravitational field into the current simple model of Callisto, thereby finalizing the full model of Callisto established in this study. Furthermore, the six parameters describing the rotation derived in this section will be utilized in the subsequent data fitting process.

\begin{figure}
\centering
\includegraphics[width=0.8\hsize]{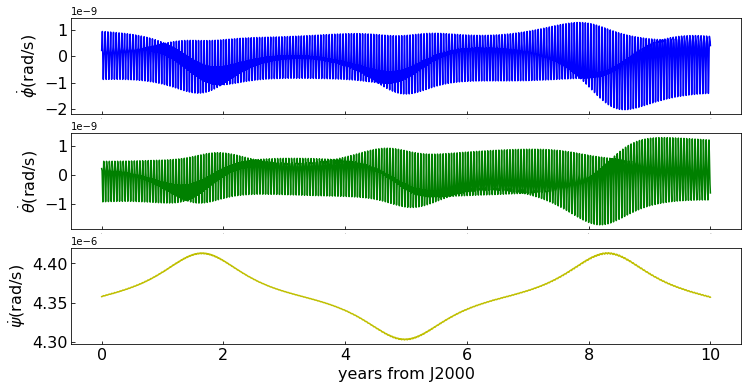}
\caption{The established rotation model of Callisto is integrated, and the temporal variations of the three Euler Angular Velocity are presented.}
\label{Figure6}
\end{figure}

\subsection{Results of the full model.}
Based on the established rotation model, we have incorporated the acceleration due to Callisto’s gravitational field as a perturbation into the satellite’s orbital motion. By replacing the librations model in Equation~\eqref{eq2} with the full rotation model of Callisto established in this paper, and by combining Equation~\eqref{eq2} with Equation~\eqref{eq18}, the rotation-orbit coupling model can be constructed. 
\begin{equation}
\begin{cases}
	\ddot{\boldsymbol{r}}=-\frac{G\left(m_0+m\right) \boldsymbol{r}}{r^3}+\sum_{j=1, j \neq i}^{\mathcal{N}} G m_j\left(\frac{\boldsymbol{r}_j-\boldsymbol{r}}{r_{i j}^3}-\frac{\boldsymbol{r}_j}{r_j^3}\right) +\\
	G\left(m_0+m\right) \nabla_i U_{\hat{\imath} \hat{0}}+\sum_{j=1, j \neq i}^{\mathcal{N}} G m_j \nabla_j U_{\hat{\jmath} \hat{0}} + \boldsymbol{a}_{rel}  \\
	\frac{d(I \boldsymbol{\omega})}{d t}+\boldsymbol{\omega} \times I \boldsymbol{\omega}=\boldsymbol{N}
\end{cases}
\end{equation}

\begin{figure}
\centering
\includegraphics[width=0.8\hsize]{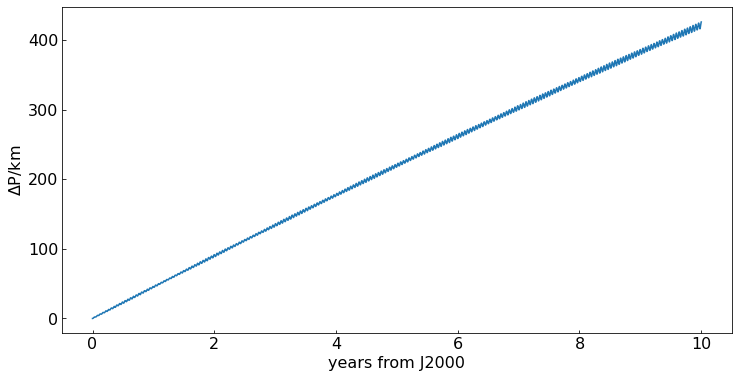}
\caption{Comparison of Callisto's orbits calculated by the full model and the simple model, where the ordinate represents positional differences in kilometers and the abscissa denotes the calculation duration in years.}
\label{diff_simple_full}
\end{figure}

In this model, the gravitational field of Callisto is truncated to second order. Starting from the J2000.0 epoch, we compared the results of integrating the full model and the simple model for a ten-year span, as shown in Figure~\ref{diff_simple_full}. The numerical experiments indicate that the differences between the two models are on the order of 400$km$. 
Among all the effects on Callisto's motion considered previously, the  rotational effects represents a significant perturbation with relatively large magnitude. 
This conclusion highlights the importance of enhancing the accuracy of both the dynamical model and the ephemerides of Callisto. The discrepancy arises from the differences in physical parameters between the simple model and the full model, as well as the introduction of Callisto’s gravitational field. It is crucial to clarify this discrepancy through the method of data fitting. 

\section{Adjustment Model}
In the previous section, we established a full model, and the results show a linear increase in discrepancy with the French NOE-5-2023 ephemerides due to the use of different physical parameters in the dynamical model. This difference can be reduced through data fitting. Therefore, in this section, we will apply the method of precise orbit determination for artificial satellites to fit the data of the full dynamical model of Callisto. In selecting the fitting data, we first use the data from the Callisto ephemerides NOE-5-2023 to fit the established simple model, with the fitted Callisto orbit data serving as the fitting data for the full model. This method ensures that the parameters in both dynamical models are consistent, and the integral result only reflects the difference caused by the inclusion of Callisto’s own gravitational field in the full model compared to the simple model.

\subsection{Variation Equations}
For a specified epoch, changes in the initial values of the state vector result in variations in the position and velocity at subsequent epochs, represented by the state transition matrix. For the full model, the variational equations are also established in accordance with this method. The difference from traditional methods lies in the expansion of the state transition matrix from 6th order to 12th order. In traditional methods, the variational equations used for precise orbit determination can be expressed as:
\begin{equation}
\frac{d(\Phi, S)}{d x}=\left(\begin{array}{cc}
	0_{3 \times 3} & 1_{3 \times 3} \\
	\frac{\partial \boldsymbol{a}}{\partial \boldsymbol{r}} & \frac{\partial \boldsymbol{a}}{\partial \boldsymbol{v}}
\end{array}\right)_{6 \times 6} \cdot(\Phi, S)+\left(\begin{array}{cc}
	0_{3 \times 6} & 0_{3 \times n_p} \\
	0_{3 \times 6} & \frac{\partial \boldsymbol{a}}{\partial p}
\end{array}\right)_{6 \times\left(6+n_p\right)}
\end{equation}
$\Phi$ represents the state transition matrix, $S$ denotes the sensitivity matrix, $p$ stands for the physical parameters to be estimated, and $n_p$ indicates the number of physical parameters to be estimated.

For the full model, involving both the rotation and orbital parameters (12 state variables in total), the structure of the state transition matrix is outlined in Equation~\ref{eq24}.
\begin{equation}
\label{eq24}
\begin{array}{|c|c|c|c|c|c|c|c|c|c|c|c|c|}
	\hline & x & y & z & v_x & v_y & v_z & \phi & \theta & \psi & \dot{\phi} & \dot{\theta} & \dot{\psi} \\
	\hline v_x & 0 & 0 & 0 & 1 & 0 & 0 & 0 & 0 & 0 & 0 & 0 & 0 \\
	\hline v_y & 0 & 0 & 0 & 0 & 1 & 0 & 0 & 0 & 0 & 0 & 0 & 0 \\
	\hline v_z & 0 & 0 & 0 & 0 & 0 & 1 & 0 & 0 & 0 & 0 & 0 & 0 \\
	\hline a_x & \frac{\partial a_x}{\partial x} & \frac{\partial a_x}{\partial y} & \frac{\partial a_x}{\partial z} & \frac{\partial a_x}{\partial v_x} & \frac{\partial a_x}{\partial v_y} & \frac{\partial a_x}{\partial v_z} & \frac{\partial a_x}{\partial \phi} & \frac{\partial a_x}{\partial \theta} & \frac{\partial a_x}{\partial \psi} & 0 & 0 & 0 \\
	\hline a_y & \frac{\partial a_y}{\partial x} & \frac{\partial a_y}{\partial y} & \frac{\partial a_y}{\partial z} & \frac{\partial a_y}{\partial v_x} & \frac{\partial a_y}{\partial v_y} & \frac{\partial a_y}{\partial v_z} & \frac{\partial a_y}{\partial \dot{\phi}} & \frac{\partial a_y}{\partial \theta} & \frac{\partial a_y}{\partial \psi} & 0 & 0 & 0 \\
	\hline a_z & \frac{\partial a_z}{\partial x} & \frac{\partial a_z}{\partial y} & \frac{\partial a_z}{\partial z} & \frac{\partial a_z}{\partial v_x} & \frac{\partial a_z}{\partial v_y} & \frac{\partial a_z}{\partial v_z} & \frac{\partial a_z}{\partial \phi} & \frac{\partial a_z}{\partial \theta} & \frac{\partial a_z}{\partial \psi} & 0 & 0 & 0 \\
	\hline \dot{\phi} & 0 & 0 & 0 & 0 & 0 & 0 & 0 & 0 & 0 & 1 & 0 & 0 \\
	\hline \dot{\theta} & 0 & 0 & 0 & 0 & 0 & 0 & 0 & 0 & 0 & 0 & 1 & 0 \\
	\hline \dot{\psi} & 0 & 0 & 0 & 0 & 0 & 0 & 0 & 0 & 0 & 0 & 0 & 1 \\
	\hline \ddot{\phi} & \frac{\partial \ddot{\phi}}{\partial x} & \frac{\partial \ddot{\phi}}{\partial y} & \frac{\partial \ddot{\phi}}{\partial z} & 0 & 0 & 0 & \frac{\partial \ddot{\phi}}{\partial \dot{\phi}} & \frac{\partial \ddot{\phi}}{\partial \theta} & \frac{\partial \ddot{\phi}}{\partial \psi} & \frac{\partial \ddot{\phi}}{\partial \dot{\phi}} & \frac{\partial \ddot{\phi}}{\partial \dot{\theta}} & \frac{\partial \ddot{\phi}}{\partial \dot{\psi}} \\
	\hline \ddot{\theta} & \frac{\partial \ddot{\theta}}{\partial x} & \frac{\partial \ddot{\theta}}{\partial y} & \frac{\partial \ddot{\theta}}{\partial z} & 0 & 0 & 0 & \frac{\partial \ddot{\theta}}{\partial \phi} & \frac{\partial \ddot{\theta}}{\partial \theta} & \frac{\partial \ddot{\theta}}{\partial \psi} & \frac{\partial \ddot{\theta}}{\partial \dot{\phi}} & \frac{\partial \ddot{\theta}}{\partial \dot{\theta}} & \frac{\partial \ddot{\theta}}{\partial \dot{\psi}} \\
	\hline \ddot{\psi} & \frac{\partial \ddot{\psi}}{\partial x} & \frac{\partial \ddot{\psi}}{\partial y} & \frac{\partial \ddot{\psi}}{\partial z} & 0 & 0 & 0 & \frac{\partial \ddot{\psi}}{\partial \phi} & \frac{\partial \ddot{\psi}}{\partial \theta} & \frac{\partial \ddot{\psi}}{\partial \psi} & \frac{\partial \ddot{\psi}}{\partial \dot{\phi}} & \frac{\partial \ddot{\psi}}{\partial \dot{\theta}} & \frac{\partial \ddot{\psi}}{\partial \dot{\psi}} \\
	\hline
\end{array}
\end{equation}
In this formula, the upper left 6-by-6 matrix of partial derivatives between orbital parameters is a commonly used state transition matrix in the precise orbit determination of artificial satellites. The lower right 6-by-6 matrix of partial derivatives between the parameters describing rotation is the state transition matrix used in the study of rotation. The two remaining parts represent the partial derivative matrices between orbital parameters and rotation parameters, as well as between rotation parameters and orbital parameters, illustrating the coupling effects between rotation and orbit in the orbit determination process. These are combined to form a 12-order state transition matrix, which is then integrated into the orbit integration for calculation. The main formula calculations in the state transition matrix refer to \citet{articleHUANG}.

\begin{figure}
\centering
\includegraphics[width=0.8\hsize]{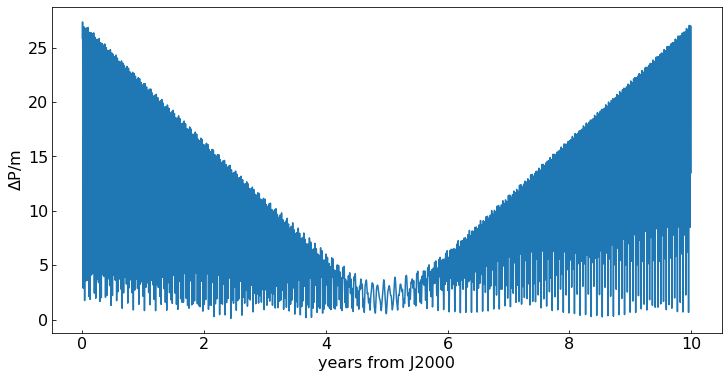}
\caption{The full model of Callisto incorporating rotation-orbit coupling is subjected to data fitting using orbital data from the reproduced simple model in Section 3. The figure presents the orbital differences between the fitted results and the simple model, where the ordinate represents positional differences in meters and the abscissa denotes the calculation duration in years.}
\label{fig:diffsimplefull}
\end{figure}

Ultimately, employing this methodology, the state transition matrix is integrated synchronously with the orbital trajectory, and the observed data is then fit using the least squares method. The computed results are contrasted with the simple model, with the discrepancies depicted in Figure~\ref{fig:diffsimplefull}.

Starting from the J2000.0 epoch, the result of the data fitting for the full dynamical model of Callisto exhibits a maximum discrepancy of approximately 25 meters with the simple model, indicating the stability and reliability of the established dynamical model and the adjustment model. 
Meanwhile, the orbital discrepancies induced by rotational effects in the full model are important for high-precision ephemeris construction of Callisto.

\section{Tidal Effect of Callisto}

In previous studies, Callisto was treated as an ideal rigid body with a single layer structure, and thus the tidal effects of Callisto were not considered. In this section, we will further study the influence of Callisto’s multi-layer structure and tidal effects in the dynamical model based on the research methods described in the previous sections.

Due to the gravitational influence of other celestial bodies in the solar system, Callisto’s shape will change under the condition of non-absolute rigidity, leading to changes in its gravitational field. For a celestial body with a mass 
$m$ and a radius $R_C$ , the solid tides generated on Callisto will have a gravitational potential function with respect to Jupiter as~\citep{Yoder2003FluidCS,Goossens2008LunarD2}:
\begin{equation}
V_{\text{tide}} = \frac{k_2}{2} \frac{G M_i}{|R_i|^3} \frac{R_C^5}{|\boldsymbol{r}|^3} \left( 3 (\hat{R}_i \cdot \hat{r})^2 - 1 \right)
\end{equation}

\begin{figure}[htbp]
\centering
\subfloat[When a fluid layer is included, $k2 = 0.4$\label{Fig2layer2}]{
	\includegraphics[width=0.6\hsize]{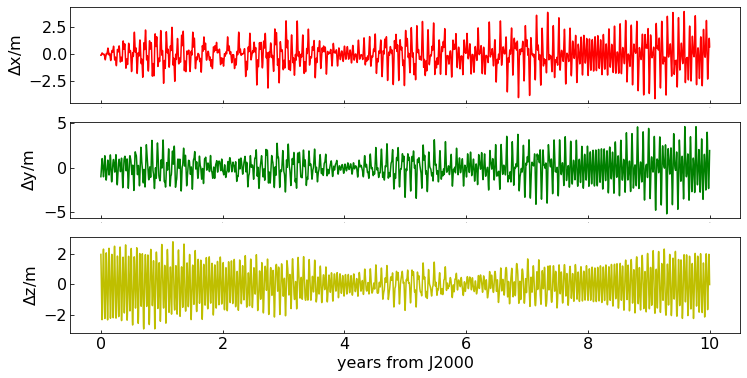}
}
\hfill
\subfloat[In the case where no fluid layer is included,$k_2 = 0.1$\label{Fig3layer2}]{
	\includegraphics[width=0.6\hsize]{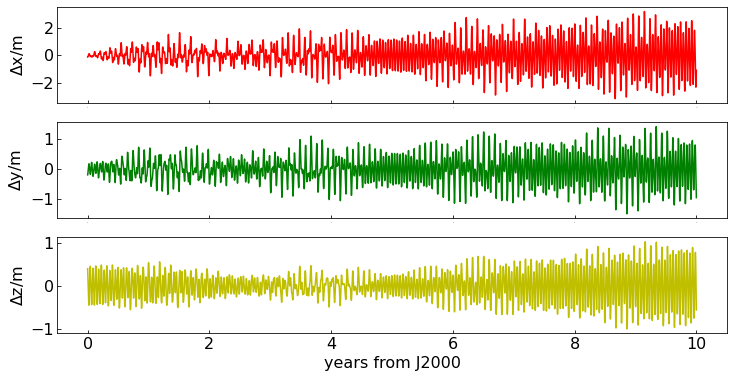}
}
\caption{The integration initiates at the J2000 epoch. The orbital data from the full model in Chapter 4 is employed to fit the dynamical model accounting for Callisto's tidal effects. The calculation results exhibit the positional differences induced by Callisto's solid tidal effects. in Figure a, Callisto has a $k_2$
	value of 0.4, and in Figure b, the $k_2$ value is 0.1. The ordinate denotes the positional differences (in kilometers), while the abscissa represents the integration duration (in years).}
\label{fig:tidefullxyz}
\end{figure}

Where $R_i$ is the position vector from the perturbing body to Callisto, 
$\boldsymbol{r}$ is the position vector from Jupiter to Callisto, $\hat{R}_i$  the unit vector from the centre of the Callisto
to the disturbing body,  $\hat{r}$ the unit vector from the centre
of the Callisto to Jupiter. 
If Callisto has an ocean, the Love number is 0.4. But if it doesn’t, the value of the Love number is approximately 0.1~\citep{Genova2022GeodeticIO,WOS:000186589600017}.
This perturbation force is added to the full model established in the previous section. The difference of the impact of Callisto’s solid tides in numerical integration between the two models is shown in Figure~\ref{fig:tidefullxyz}. The results indicate that the tidal effects on Callisto can cause a difference of several meters over ten years of orbit.
%\begin{figure}
%	\centering
%	\includegraphics[width=0.8]{TIDE}
%	\caption{The effect of Callisto’s tides on its orbital motion.}
%	\label{tide}
%\end{figure}

\begin{figure}[htbp]
\centering
\subfloat[The double-layer structure model of Callisto\label{Fig2layer}]{
	\includegraphics[width=0.4\hsize]{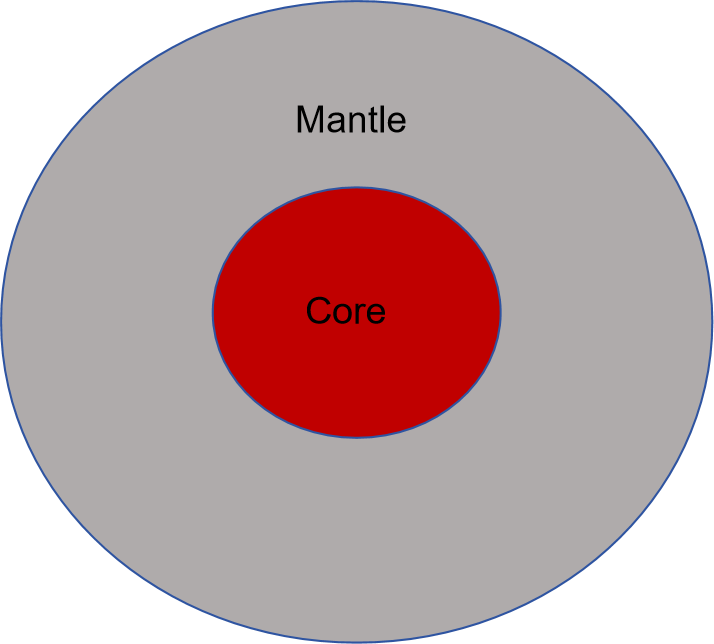}
}
\hfill
\subfloat[The 3-layer structure model of Callisto\label{Fig3layer}]{
	\includegraphics[width=0.4\hsize]{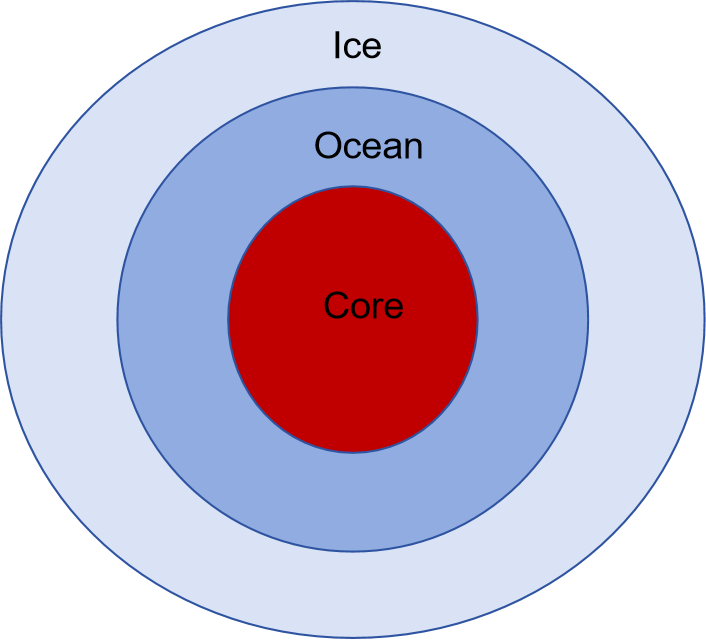}
}
\caption{The internal structure model of Callisto employed in current mainstream research.}
\end{figure}

We consider two different internal structures of Callisto based on some existing studies on its internal structure~\citep{Anderson2001ShapeMR,Nagel2003AMF,Genova2022GeodeticIO}: (1) A thick (more than about 1000 kilometers) ice and rock-metal crust, with a density close to 1600 $kg/m^3$, covering a rocky-metallic core with a density close to 2300 $kg/m^3$. As the Figure~\ref{Fig2layer} shown. (2) A three-layer structure of ice-water-core, featuring a relatively pure ice shell, approximately 300 kilometers thick with a density of 1050 $kg/m^3$, covering an internal ocean with a thickness of approximately 20-200 $km$, and a solid core as the innermost layer. As the Figure~\ref{Fig3layer} shown.
%\begin{figure}
%	\centering
%	\includegraphics[width=0.5\hsize]{2 layer}
%	\caption{The double-layer structure model of Callisto.}
%	\label{Fig2layer}
%\end{figure}
%\begin{figure}
%	\centering
%	\includegraphics[width=0.5\hsize]{3 layer}
%	\caption{The 3-layer structure model of Callisto.}
%	\label{Fig3layer}
%\end{figure}

(1) For the first internal structure model, we refer to the research methods on the layered structure of the Moon to calculate the moment of inertia of Callisto~\citep{Folkner2014ThePA,Pavlov2016DeterminingPO}. The moment of inertia of the core can be expressed as:
\begin{equation}
\mathbf{I}_c=\left[\begin{array}{ccc}
	A_c & 0 & 0 \\
	0 & B_c & 0 \\
	0 & 0 & C_c
\end{array}\right].
\end{equation}

where $A_c, B_c$, and $C_c$ are the moments of inertia about the three principal axes of the core when there is no deformation. It is assumed that during the motion, the core does not undergo deformation, and the moment of inertia of the undistorted core is the difference between the total undistorted moment of inertia and the moment of inertia of the core.
\begin{equation}
\mathbf{I}_m = \mathbf{I}_T - \mathbf{I}_c.
\end{equation}

The tidal deformation of Callisto 
due to Jupiter and other celestial bodies results in a change in the moment of inertia of Callisto’s mantle over time. The moment of inertia of Callisto’s mantle can be expressed as:

\begin{equation}
\begin{aligned}
	\mathbf{I}_m(t)= & \tilde{\mathbf{I}}_m-\frac{k_{2} m R_C^5}{r^5}\left[\begin{array}{ccc}
		x^2-\frac{1}{3} r^2 & x y & x z \\
		x y & y^2-\frac{1}{3} r^2 & y z \\
		x z & y z & z^2-\frac{1}{3} r^2
	\end{array}\right] \\
	& +\frac{k_{2} R_C^5}{3 G}\left[\begin{array}{ccc}
		\omega_{x}^2-\frac{1}{3}\left(\omega^2-n^2\right) & \omega_{x} \omega_{y} & \omega_{x} \omega_{z} \\
		\omega_{x} \omega_{y} & \omega_{y}^2-\frac{1}{3}\left(\omega^2-n^2\right) & \omega_{y} \omega_{z} \\
		\omega_{x} \omega_{z} & \omega_{y} \omega_{z} & \omega_{z}^2-\frac{1}{3}\left(\omega^2+2 n^2\right)
	\end{array}\right].
\end{aligned}
\end{equation}

$k2$ is the Callisto potential Love number; $M$ is the mass of 
the Jupiter; $R_C$ is the equatorial radius of the Callisto; $r$ is the Jupiter-Callisto distance;

\begin{figure}
\centering
\includegraphics[width=0.8\hsize]{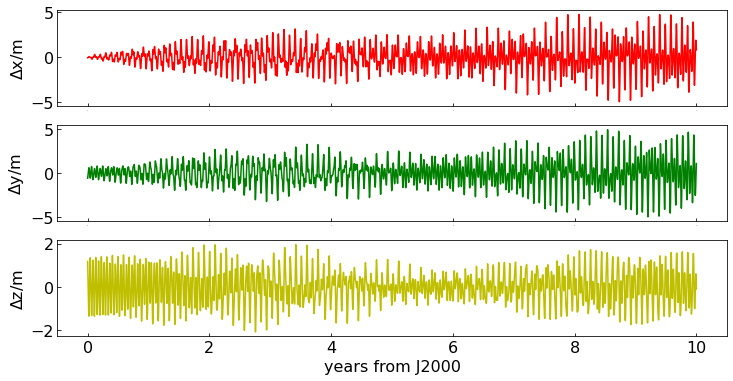}
\caption{The orbital difference between the Callisto two-layer structure model and the single-layer model(after data fitting).The fitting method employs the fitted orbital data of Callisto's full model, calculated in Chapter 4, to perform data fitting on Callisto's two-layer structure model. The ordinate denotes the positional differences (in meters), while the abscissa represents the integration duration (in years).}
\label{fig:twolayerfit}
\end{figure}

When the moment of inertia of the mantle changes, Callisto’s rotation will also change accordingly. By incorporating the new rotation tensor matrix into the model established in the previous section and using the data fitting of the full model as the fitting data for the new model, we ensure that the results only reflect the differences due to the different internal structure models of Callisto. The numerical results are shown in Figure~\ref{fig:twolayerfit}, and the results indicate that the difference over ten years is approximately 5 $m$.

(2) Now, considering the second internal structure model, it is generally believed that Callisto has an underground ocean, forming a three-layer structure of ice-ocean-rock. Under this structure, we consider the viscous friction between the middle layer of water and the ice layer as well as the rock layer. The outermost layer of Callisto is ice with a density of 1050 $kg/m^3$ and a thickness of approximately 300 km. Assuming the thickness of the internal ocean is 200 km. In this case, the calculated Reynolds number for the fluid motion is approximately $10^{11}$, indicating that the liquid layer is primarily moving in a turbulent state. The Fanning equation can be used to calculate the energy loss per unit mass:
\begin{equation}
\label{eq30}
\mathrm{h_f}=\lambda \frac{L}{d} \frac{u^2}{2}.
\end{equation}

\begin{figure}
\centering
\includegraphics[width=0.8\hsize]{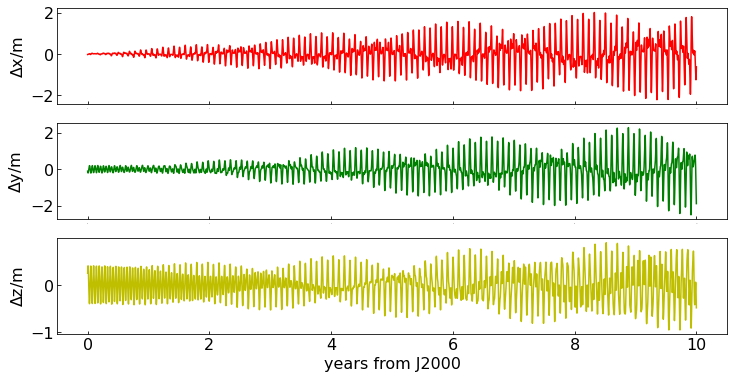}
\caption{The orbital difference between the Callisto there-layer structure model and the single-layer model(after data fitting). The fitting method employs the fitted orbital data of Callisto's full model, to perform data fitting on Callisto's there-layer structure model. The ordinate denotes the positional differences (in meters), while the abscissa represents the integration duration (in years).}
\label{fig:threelayerfitxyz}
\end{figure}

In this study, it is postulated that Callisto’s three-layered structure rotates at a uniform angular velocity. The energy loss can be calculated by Formula~(\ref{eq30}). and the equivalent torque of this energy loss is substituted into Formula~(\ref{eq22}). Subsequently, in accordance with the model established in the preceding sections and the data fitting methodology employed, the moment of inertia of Callisto is the sum of the moment of inertia of the ice shell and the moment of inertia of the internal spherical core. After numerical integration, the alterations in Callisto’s orbit due to the incorporation of the new torques are depicted in Figure~\ref{fig:threelayerfitxyz}. The results indicate that over a decade of orbital evolution, this structure will lead to a difference of meters.

In summary, for the two-layer model, we considered the difference in moment of inertia caused by the different densities between the core and the mantle, and then calculated the orbital changes of Callisto resulting from moment of inertia using the model established in this paper. For the three-layer model, this paper analyzed the viscous torque induced by viscous forces and obtained the influence of this torque on the orbit. The conclusions will facilitate our in-depth analysis of Callisto's internal structure using the high-precision observation data from the future Tianwen-4 mission.

\section{Conclusions}

This study introduces a new dynamical model of Callisto, which, building upon existing models, fully incorporates Callisto’s rotation, thereby establishing a dynamical model coupling Callisto’s rotation and orbital motion. Furthermore, based on the full model, the orbital motion discrepancies under several mainstream internal structural models of Callisto are calculated. The establishment of this model provides a reference orbit for Callisto for the upcoming Tianwen-4 mission, while also serving as a testing platform for the data to be collected by the “Tianwen-4” probe.

In the process of data fitting, a variational equation for fitting the full model was established, and the analytical expressions for the coupled effects of Callisto’ rotation and orbit were solved. Starting from the J2000.0 epoch, the variational equation was integrated over a period of ten years. The final results indicate that the difference from the  ephemeris NOE-5-2023 is no more than 100 meters, which fully demonstrates its applicability.

With the advancement of various exploration missions targeting Callisto, a substantial amount of high-precision data for Callisto is anticipated. Utilizing this data, in conjunction with the full dynamical model established in this study, there is potential to develop a new generation of ephemerides for Callisto with enhanced accuracy. Additionally, the methods presented in this paper can be employed to constrain the internal structure of Callisto.

The full model offers a new alternative for the construction of a new ephemeris for Callisto. Furthermore, future high-precision missions targeting Callisto—such as Tianwen-4 and JUICE—will yield more high-accuracy data. This, in turn, will effectively refine Callisto's current physical parameters, including its gravity field model and a more precise $K_2$. This is of great significance to the present study, as it will further enhance the full model's accuracy and enable deeper research into internal structures using the model. Additionally, The full model can provide a reference dynamical model for the development of numerical ephemerides for the major bodies of the solar system for China.

%% Please use the acknowledgment and contribution environments. This will 
%% be anonomyized when the "anonymous" style option is used. 
\begin{acknowledgments}
This research was supported by the Strategic Priority Research Program of the Chinese Academy of Sciences (XDA0350300), the National Key Research and Development Program of China (2021YFA0715101), the National Natural Science Foundation of China (12033009, 12103087),  the International Partnership Program of Chinese Academy of Sciences (020GJHZ2022034FN), the Yunnan Fundamental Research Projects (202201AU070225, 202301AT070328, 202401AT070141), the Young Talent Project of Yunnan Revitalization Talent Support Program. We would like to thank Professor Valery Lainey for his guidance on the data fitting method. We thank the anonymous referee for their careful reviewing and nice suggestions that have been incorporated into and improved the paper. 
\end{acknowledgments}

\begin{contribution}
%%This section gives authors the space to recognize author contributions. The text inside this environment is NOT counted towards the total word quanta. At a minimum, manuscripts are expected to include this text:

All authors contributed equally to this paper.

%% But authors are expected to provide more specific details, e.g. 
%%
%%SC was responsible for writing and submitting the manuscript.
%%WWM came up with the initial research concept and edited the manuscript.
%%OTS obtained the funding and edited the manuscript.
%%EBF provided the formal analysis and validation. He also edited the manuscript.
%%GEH Supervised the undergraduates, wrote the software and administers the project github and Zenodo repositories.
%%
%% Authors can use the Contributor Role Taxonomy (CRediT) at
%% for ideas on how write a good statement tailored to their needs.

\end{contribution}

\bibliography{sample7}{}
\bibliographystyle{aasjournal}



\end{document}